\documentclass[12pt,preprint]{emulateapj}
\usepackage{epsfig}
\usepackage{graphicx}
\shorttitle{}
\shortauthors{Dow-Hygelund et al.}
\begin{document}
\title{UV Continuum Spectroscopy of a 6$L_{*}$ \textit{z} = 5.5 Starburst Galaxy \altaffilmark{1,}\altaffilmark{2,}\altaffilmark{3}}

\author{C.C. Dow-Hygelund\altaffilmark{4}, B.P. Holden\altaffilmark{5}, R.J. Bouwens\altaffilmark{5},  A. van der Wel\altaffilmark{6}, G.D. Illingworth\altaffilmark{5}, A. Zirm\altaffilmark{6}, M. Franx\altaffilmark{6}, P. Rosati\altaffilmark{7}, H. Ford\altaffilmark{8}, P.G. van Dokkum\altaffilmark{9}, S.A. Stanford\altaffilmark{10},
  P. Eisenhardt\altaffilmark{11}, and G.G. Fazio\altaffilmark{12}}
\altaffiltext{1}{Based on observations made with the NASA/ESA \textit{Hubble Space Telescope}, which is operated by the Association of Universities for Research in Astronomy, Inc., under NASA contract NAS 5-26555.  These observations are associated with programs 7817, 9270, 9301, and 9583}
\altaffiltext{2}{Based on observations collected at the European Southern Observatory, Paranal, Chile (LP166.A-0701 and 169.A-0458).}
\altaffiltext{3}{This work is based in part on observations made with the \textit{Spitzer Space Telescope}, which is operated by the Jet Propulsion Laboratory,
California Institute of Technology under NASA contract 1407.}
\altaffiltext{4}{Physics Department, University of California,
    Santa Cruz, CA 96062; cdow@scipp.ucsc.edu}
\altaffiltext{5}{Astronomy Department, University of California,
    Santa Cruz, CA 96062; holden@ucolick.org, bouwens@ucolick.org, gdi@ucolick.org}
\altaffiltext{6}{Leiden Observatory, P.O. Box 9513, J. H. Oort Building, Niels Bohrweg 2, NL-2300 RA, Leiden, Netherlands; vdwel@strw.leidenuniv.nl, franx@strw.leidenuniv.nl, azirm@strw.leidenuniv.nl}
\altaffiltext{7}{European Southern Observatory,
  Karl-Schwarzchild-Strasse 2, D-85748 Garching, Germany;
  prosati@eso.org}
\altaffiltext{8}{Department of Physics and Astronomy, Johns Hopkins University,  3400 North Charles Street
Baltimore, MD 21218-2686; ford@pha.jhu.edu}
\altaffiltext{9}{Department of Astronomy, Yale University, P.O. Box 208101, New Haven, CT 06520-8101; dokkum@astro.yale.edu}
\altaffiltext{10}{Department of Physics, University of California, Davis}
\altaffiltext{11}{Jet Propulsion Laboratory, California Institute of Technology, MS 169-327, 4800 Oak Grove Drive, Pasadena, CA 91109}
\altaffiltext{12}{Harvard-Smithsonian Center for Astrophysics, 60 Garden Street, MS 65, Cambridge, MA 02138}
\begin{abstract}

We have obtained a high S/N (22.3 hr integration) UV continuum VLT
FORS2 spectrum of an extremely bright ($z_{850}$ = 24.3) \textit{z} =
5.515 $\pm$ 0.003 starforming galaxy (BD38) in the field of the
\textit{z} = 1.24 cluster RDCS 1252.9-2927.  From \textit{HST}
Advanced Camera for Surveys imaging this object was selected as a
potential \textit{z} $\sim$ 6 Lyman break galaxy (LBG) based on its
red $i_{775}-z_{850}$ = 1.5 color.  This object shows substantial
continuum (0.41 $\pm$ 0.02 $\mu$Jy at $\lambda$1300) and
low-ionization interstellar absorption features typical of LBGs at
lower redshift (\textit{z} $\sim$ 3); this is the highest redshift LBG
confirmed via metal absorption spectral features.  The equivalent
widths of the absorption features are similar to \textit{z} $\sim$ 3
strong Ly$\alpha$ absorbers.  No noticeable Ly$\alpha$ emission was
detected ($F$ $\leq$ $1.4$ $\times$ $10^{-18}$ ergs cm$^{-2}$
s$^{-1}$, 3$\sigma$).  This object is at most amplified 0.3 mag from
gravitational lensing by the foreground cluster.  The half-light
radius of this object is 1.6 kpc (0\farcs25) and the star formation
rate derived from the rest-frame UV luminosity is $SFR_{UV}$ = 38
$h^{-2}_{0.7} \ M_{\sun} \ \rm{yr^{-1}}$ (142 $h^{-2}_{0.7} \ M_{\sun}
\ \rm{yr^{-1}}$ corrected for dust extinction).  In terms of recent
determinations of the \textit{z} $\sim$ 6 UV luminosity function, this
object appears to be 6$L_{*}$.  The \textit{Spitzer} IRAC fluxes for
this object are 23.3 and 23.2 AB mag (corrected for 0.3 mag of cluster
lensing) in the 3.6$\mu$ and 4.5$\mu$ channels, respectively, implying
a mass of 1-6 $\times$ 10$^{10}$ $M_{\sun}$ from population synthesis
models.  This galaxy is brighter than any confirmed \textit{z} $\sim$
6 \textit{i}-dropout to date in the $z_{850}$ band, and both the
3.6$\mu$ and 4.5$\mu$ channels, and is the most massive starbursting
galaxy known at \textit{z} $>$ 5.

\end{abstract}


\section{Introduction}

Despite the large samples of photometrically selected \textit{z}
$\sim$ 6 Lyman Break Galaxies (LBGs) \citep{Bouwens2003,Bouwens2005},
spectroscopic confirmation has been difficult, because of the faint
flux levels required.  Ground based follow-up programs have had
success rates of $\sim$25$\%$ by identifying redshifts through
Ly$\alpha$ emission
\citep{Stanway2004b,Stanway2004a,DowHygelund2005a}.
\citet{Malhotra2005} have verified 23 \textit{z} $\sim$ 6 $i_{775}$
dropout galaxies via the presence of strong Lyman continuum breaks
using low resolution ACS Grism observations.  However, none of these
very low resolution spectra can be used to identify the absorption
features necessary to measure metal abundances and to characterize the
interstellar medium.  To do this we need to observe anomalously bright
objects to obtain high continuum S/N.  The best example of this
remains the strongly gravitationally lensed (30$\times$) \textit{z}
$\sim$ 3 galaxy MS 1512-cB58 \citep{Pettini2000}.

In this Letter, we report on the spectroscopic confirmation of a
particularly bright $z_{850}$ = 24.3 \citet{Bouwens2003} \textit{z}
$\sim$ 6 \textit{i}-dropout candidate: object 1252-5224-4599,
hereafter BD38, in the RDCS 1252.9-2927 field (CL1252)
\citep{Rosati1998}.  We show that this large ($r_{hl}$ = 0\farcs29)
starbursting galaxy has a Lyman continuum break and interstellar
absorption features at a redshift of \textit{z} = 5.515, but lacks
Ly$\alpha$ emission.  Throughout we adopt ($\Omega_{tot}$,
$\Omega_{M}$, $\Omega_{\Lambda}$) = (1.0, 0.3, 0.7) and $H_{o}$ = 70
km s$^{-1}$ Mpc$^{-1}$.  All magnitudes are given in the AB system
\citep{Oke&Gunn1983}.

\section{Observations, Reduction, and Analysis}
The \textit{Hubble Space Telescope} Advanced Camera for Surveys (ACS)
observations of BD38 ($\alpha_{2000} = 12^h52^m56\fs888, \delta_{2000}
= -29\degr25\arcmin55\farcs50$) involved three orbits of F775W
($i_{775}$) and five orbits of F850LP ($z_{850}$).  VLT ISAAC (6.0 hr
for \textit{Js} and 5.7 hr for \textit{Ks}) and \textit{Spitzer} IRAC
(1000 s for both 3.6$\mu$ and 4.5$\mu$) imaging were also obtained.
For details see \citet{Bouwens2003}, \citet{Lidman2004}, and
\citet{Fazio2004}.

We measured the $i_{775} - z_{850}$ color magnitudes in 0\farcs6
apertures, and the total $z_{850}$ magnitude was measured in a
1\farcs1 aperture.  For measuring the $z_{850}-Js$ and $z_{850}-Ks$
colors, we smoothed the $z_{850}$ image to match the \textit{Js}- and
\textit{Ks}-band psf, and used 1\farcs0 apertures.  The difference
between the 1\farcs1 and 0\farcs6 $z_{850}$ magnitudes was then added
to the $i_{775}$, \textit{Js}, and \textit{Ks} magnitudes.

Using the $z_{850}$ image as a model for the IRAC data, every object in
the $z_{850}$ image within 20\arcsec\ of BD38 was smoothed with a
kernel to match the ACS and IRAC data.  The normalization of
this kernel was varied individually for every object until the
$\chi^{2}$ between the model image and the IRAC data was minimized.

\begin{figure*}
\begin{center}
\includegraphics[width=6.8in, angle=0]{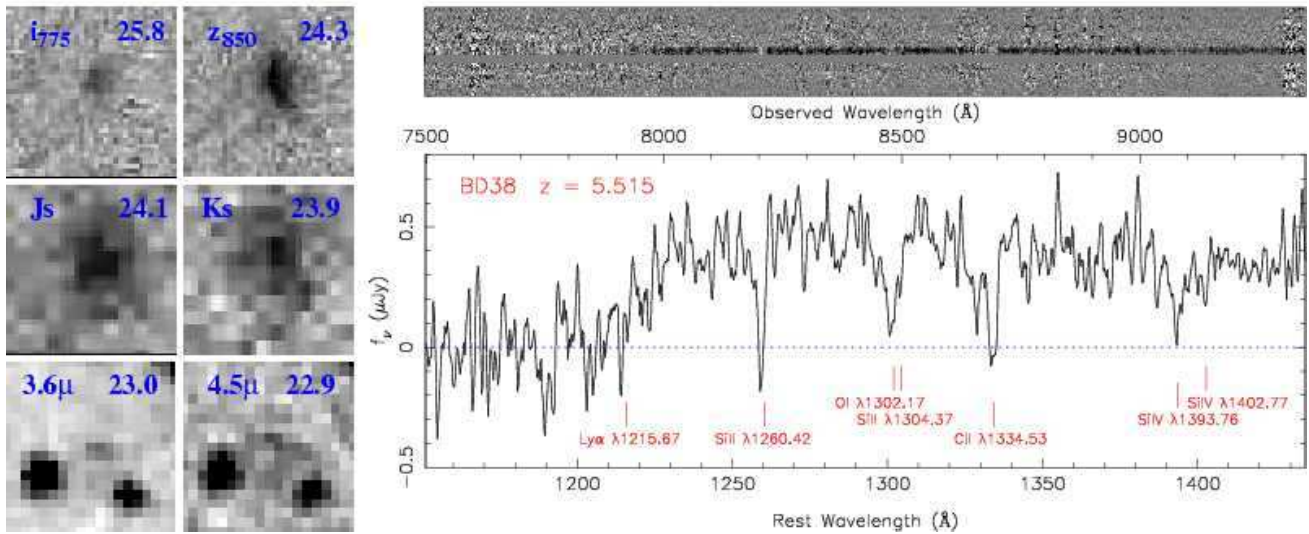}
\caption{\em (Left)  ACS, ISAAC, and IRAC imaging of BD38.  The
  $i_{775}$, $z_{850}$, $Js$, $Ks$ thumbnails are 1\farcs5 on a side.
  The 3.6$\mu$ and 4.5$\mu$ thumbnails are 15\farcs0 on a side.  The
  disturbed morphology of BD38 is apparent and, together with the
  spectroscopic data, suggests vigorous star formation.  Furthermore,
  the relatively red \textit{z-Js} and \textit{z-Ks} colors coupled
  with the strong LIS lines (see below), suggests BD38 is a relatively
  dust enriched and chemically evolved galaxy.  The object is extended
  towards the lower-right
  (north-west) direction by $\approx$ 0\farcs26 and by 0\farcs22
  upward (east).  These ``plumes'' could be the product of a recent
  merger.  (Right)  Two- and extracted one-dimensional FORS2 spectra of BD38.
  The total integrated time was 22.3 hr.  The gray smooth rectangular
  region of the two-dimensional spectrum covers a cluster member
  spectrum that occupied the slit.  The one-dimensional spectrum has
  been smoothed with a 5 pixel boxcar filter.  Strong LIS absorption
  features are marked, and the blue dotted
  line denotes zero flux.  The equivalent widths of these lines
  are very similar to those found by \citet{Shapley2003} for 198
  \textit{z} $\sim$ 3 LBGs that are strong Ly$\alpha$ absorbers (see
  Table \ref{BD38_spectral}).  Utilizing these features yields a systematic redshift of \textit{z} = 5.515 $\pm$ 0.003.
\label{4202}}
\end{center}
\end{figure*}

The resulting ACS, ISAAC and IRAC images are shown in Figure
\ref{4202}.  BD38's observed magnitudes are $i_{775}$ = 25.76 $\pm$
0.15, $z_{850}$ = 24.25 $\pm$ 0.05, \textit{Js} = 24.1 $\pm$ 0.1,
\textit{Ks} 23.8 $\pm$ 0.1, 3.6$\mu$ = 23.0 $\pm$ 0.3, and 4.5$\mu$ =
22.9 $\pm$ 0.4.  Clearly evident is the strong $i_{775}-z_{850}$ flux
decrement.  Furthermore, the object is increasing in luminosity
towards longer wavelengths.  Assuming the rest-frame UV continuum of
BD38 satisfies the form $f_{\lambda}\propto\lambda^{\beta}$ we are
able to derive a spectral slope of $\beta$ = -1.5 $\pm$ 0.2 over rest
frame $\lambda$1400-$\lambda$3000, from the $z_{850}$, \textit{Js},
and \textit{Ks} fluxes.

We spectroscopically observed BD38 using the Focal Reducer/low
dispersion Spectrograph 2 (FORS2) on the 8.2-m VLT YEPUN Unit
Telescope.  We used the 600z grism with the OG590 blocking filter,
yielding a resolution of 1.64 \AA\ per pixel.  The slit width was
1\farcs0.  A total of 80 exposures were taken to acquire a 22.3 hr
integrated spectrum.  BD38 and three other \textit{i}-dropouts were
targeted serendipitously as part of a larger fundamental plane study.
For details see \citet{Arjen2005}.
 
The 22.3 hr integrated FORS2 absorption spectra are shown in Figure
\ref{4202}.  A precipitous continuum break is clearly seen at
$\approx$\ 7900 \AA, reducing the continuum from 0.41 $\pm$ 0.02 to 0
$\pm$ 0.02 $\mu$Jy.  Six clear absorption features are indicated in
the figure.

\begin{figure}
\begin{center}
\includegraphics[width=3.5in, angle=270, scale=0.9]{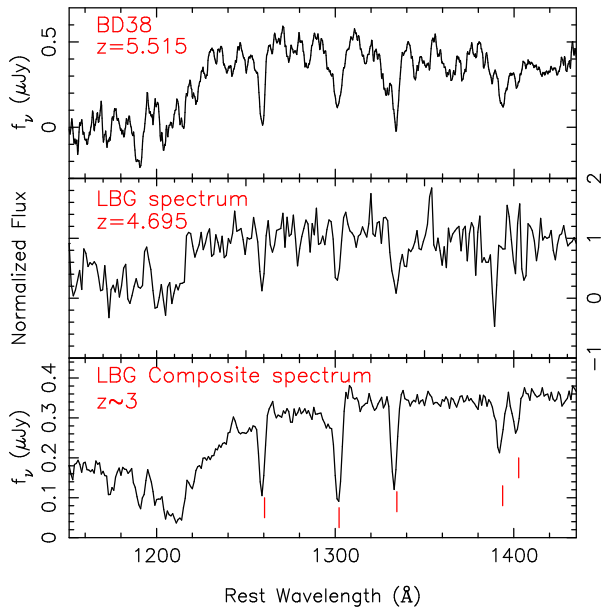}
\caption{\em Spectral comparison between BD38, a \textit{z} = 4.695 LBG spectrum from \citet{Ando2004}, and \citet{Shapley2003} \textit{z} $\sim$ 3 composite LBG spectrum.  BD38's spectrum has been smoothed with a 13 pixel boxcar filter.  The continuum break and marked absorption features show a strong match with BD38 and the LBG spectra, indicating this object is a starbursting LBG.
\label{4202_comp}}
\end{center}
\end{figure}

The continuum break is due to the Ly$\alpha$ forest, while the
absorption features are the low-ionization interstellar (LIS) lines
typically seen in LBGs at \textit{z} $\sim$ 3 \citep{Shapley2003}, and
the UV spectra of local starbursts \citep[e.g.,][]{Heckman1998}.  In
Figure \ref{4202_comp} we show BD38's spectrum along with the
composite spectrum of 198 \textit{z} $\sim$ 3 LBGs that lack
Ly$\alpha$ emission \citep{Shapley2003}, and object 2 (\textit{z} =
4.687 LBG) of \citet{Ando2004}.  There is a strong similarity in each
spectrum's continuum break and absorption features.  This demonstrates
that BD38 is a LBG at \textit{z} = 5.515 $\pm$ 0.003.

There appears to be some weak emission at 7920 $\pm$ 1 \AA\, which
would be the expected position of the Ly$\alpha$ emission line.
However, the S/N for this feature is poor.  We determine a 3$\sigma$
upper limit of $1.4$ $\times$ $10^{-18}$ ergs cm$^{-2}$ s$^{-1}$ on
this features.  This is an order of magnitude smaller than is seen for
\textit{z} $\sim$ 6 LBGs with strong Ly$\alpha$ emission
\citep{Stanway2004a,DowHygelund2005a}.

\begin{deluxetable}{cccccc}
\tabletypesize{\scriptsize}
\tablecaption{Strong BD38 Interstellar Absorption Features\label{BD38_spectral}}
\tablewidth{0pt}
\tablehead{
\colhead{$\lambda_{lab}$\tablenotemark{a}} & \colhead{Ion} & \colhead{$\lambda_{obs}$\tablenotemark{b}} & \colhead{$z_{ion}$\tablenotemark{b}} & \colhead{W$_{o}$\tablenotemark{c}} & \colhead{W$_{o,z \sim 3}$\tablenotemark{d}}\\
\colhead{(\AA)} &\colhead{} &\colhead{(\AA)} &\colhead{} &\colhead{(\AA)} &\colhead{(\AA)}
}
\startdata
1260.42  & Si \tiny{II} & 8205.77 & 5.510 & 3.18$\pm$0.15 &  1.85$\pm$0.16  \\
1302.17 & O \tiny{I}   & 8485.38 & 5.516 & 3.05$\pm$0.10 &  3.24$\pm$0.16 \\
1304.37 & Si \tiny{II} & 8498.43 & 5.515 &     -         &   -            \\
1334.53 & C \tiny{II} & 8699.72 & 5.518 & 2.88$\pm$0.13 &  2.34$\pm$0.16 \\ 
1393.76 & Si \tiny{IV} &  9079.21 & 5.514 & 1.92$\pm$0.25 &  1.83$\pm$0.23 \\
1402.77 & Si \tiny{IV} & 9140.84 & 5.516 & 1.14$\pm$0.07 &  1.01$\pm$0.17 \\
mean & & & 5.515 & & 
\enddata
\tablecomments{For ion O \tiny{I} \scriptsize, W$_{o}$ and W$_{o,z \sim 3}$ contain O \tiny{I} \scriptsize $\lambda$1302.17 + Si \tiny{II} \scriptsize $\lambda$1304.37.}
\tablenotetext{a}{Vacuum wavelengths}
\tablenotetext{b}{Observed wavelenghts (heliocentric)}
\tablenotetext{c}{Rest frame equivalent width and 1$\sigma$ error}
\tablenotetext{d}{Rest frame equivalent width for group 1 (G1) of the \citet{Shapley2003} \textit{z} $\sim$ 3 LBG sample}
\end{deluxetable}

\section{A $\sim6L_{*}$ Starburst Galaxy}

\subsection{Luminosity}

BD38 is 1.3 mag ($z_{850}$) brighter than any other \textit{i}-dropout
found in its host field CL1252 \citep{Bouwens2003}, and is 0.4 mag
brighter than any other verified \textit{z} $\sim$ 6 LBG to date
\citep{Bunker2003,Malhotra2005,DowHygelund2005a}.  Two common
explanations for anomalously bright galaxies are (1) they are powered
by an AGN or (2) they are gravitationally lensed.  We consider each
hypothesis in turn.

The AGN interpretation of the high luminosity is unlikely.  First,
Ly$\alpha$ $\lambda$1215.67 emission and N \scriptsize{V}\normalsize
$\lambda$1240 emission, typical of AGNs \citep{Osterbrock1989}, were
not detected.  Second, BD38 was undetected in Chandra and XMM Newton
CL1252 surveys, with upper luminosity limits of $L_{x}$ = 1.6 $\times$
10$^{43}$ ergs s$^{-1}$ and 6.5 $\times$ 10$^{43}$ ergs s$^{-1}$ in
the 0.5-2 keV and 2-7 keV bands, respectively \citep{Rosati2004}.
This excludes BD38 being a type-1 AGN (QSO), and the absence of high
ionization spectral lines is inconsistent with this object being a low
luminosity type-2 AGN.  Finally, the object is very extended, lacks a
point source, and has LIS features similar to lower redshift starburst
spectra.


Gravitational amplification can only account for a small portion of
BD38's high luminosity.  BD38 is 88\arcsec\ from the CL1252 cluster
center, and using the results of \citet{Lombardi2005} we estimate the
gravitational magnification due to the cluster potential to be at most
1.3$\times$ or 0.3 mag.  Lensing by any obvious nearby galaxy has been
calculated to be less than 3$\%$.  Therefore, we believe BD38 to be
intrinsically very luminous.

BD38 is an extremely luminous ($z_{850}$ = 24.6 mag delensed) galaxy
for this epoch.  If no evolution is assumed in the UV luminosity
function from \textit{z} $\sim$ 3 to \textit{z} $\sim$ 6 for $L_{*}$,
BD38 is a 4$L_{*}$ object.  However, using the observed evolution in
$L_{*}$, this object is an even more extreme 6$L_{*}$
\citep{Bouwens2005}.  The surface density of such objects is estimated
to be only one per 400 arcmin$^{2}$.  In fact, only one similarly
bright (SBM03$\#$3; $z_{850}$ = 24.7) \textit{i}-dropout has been
found over the entire 320 arcmin$^{2}$ ACS footprint of the two GOODS
fields \citep{Bunker2003}.  Furthermore, BD38 is brighter than all the
\textit{i}-dropout objects discovered in the 767 arcmin$^2$ Subaru
Deep Field \citep{Shimasaku2005}.  In addition, BD38's 3.6$\mu$ = 23.3
and 4.5$\mu$ = 23.2 mag fluxes (delensed) are the largest found in any
\textit{i}-dropout object to date \citep{Eyles2005}, and are 0.6 mag
brighter than SBM03$\#$3 in the 3.6$\mu$ channel.

The UV magnitudes imply an $\lambda$1500 continuum luminosity of 2.7
$\times$ $10^{29} \ h^{-2}_{0.7} \ \rm{ergs \ s^{-1} \ Hz^{-1}}$.
This yields a star formation rate ($SFR_{UV}$) of 38 $h^{-2}_{0.7} \
M_{\sun} \ \rm{yr^{-1}}$ \citep{Madau1998}.  The $\beta$ = -1.5
suggests that BD38 suffers from dust obscuration, and utilizing Eqs
(11) of \citet{Meurer1997} yields a UV dust extinction value
$A_{1600}$ = 1.45 magnitudes.  This increases $SFR_{UV}$ by a factor
of 3.78 to 142 $h^{-2}_{0.7} \ M_{\sun} \ \rm{yr^{-1}}$.

\subsection{Morphology and Color}

BD38 has a delensed half-light radius of $r_{hl}$ = 0\farcs25
($z_{850}$), implying a physical half light radius of $r_{hl}$ = 1.6
$h_{0.7}$ kpc.  This object is among the largest \textit{i}-dropouts
at \textit{z} $\sim$ 6.  This $r_{hl}$ corresponds to the size of
Luminous Blue Compact Galaxies in the local universe \citep{Koo1994},
though the $SFR$ of BD38 is a factor of $\approx$6$\times$ higher.

Though BD38 is unusually bright, its $A_{1600}$, $SFR_{UV}$, and
$r_{hl}$ resemble \textit{z} $\sim$ 3 LBGs and compact ultra-violet
luminous galaxies (UVLGs) populating the local universe
\citep{Heckman2005}.  Averaging the equivalent width values of
SiII(1260), OI(1302)/SiII(1304), and CiII(1335) yields -3.0 \AA,
similar to -2.5 \AA\ for G1 and -2.8 \AA\ for seven \textit{z} $\sim$
5 LBGs \citep{Ando2004}, and those found for UVLGs.  These lines are
saturated, therefore they are not useful for measuring metal
abundances \citep[see][]{Shapley2003}.  However, this similarity in the
LIS equivalent widths suggests BD38's combination of neutral
gas covering fraction and velocity dispersion is similar to lower
redshift starbursts.


\subsection{Population Synthesis Modeling}

Using the $z_{850}$, \textit{Js}, \textit{Ks}, and \textit{Spitzer}
IRAC 3.6$\mu$ and 4.5$\mu$ data, we fit \citet{Bruzual2003} stellar
sythesis models to BD38 using a similar methodology as
\citet{Papovich2001} and \citet{Eyles2005}.  We use the dust models
from \citet{Calzetti2000}, and assume metallicities
of 0.4 or 1.0 $Z_{\sun}$. 

\begin{figure}
\begin{center}
\includegraphics[width=3.4in, angle=0,scale=1]{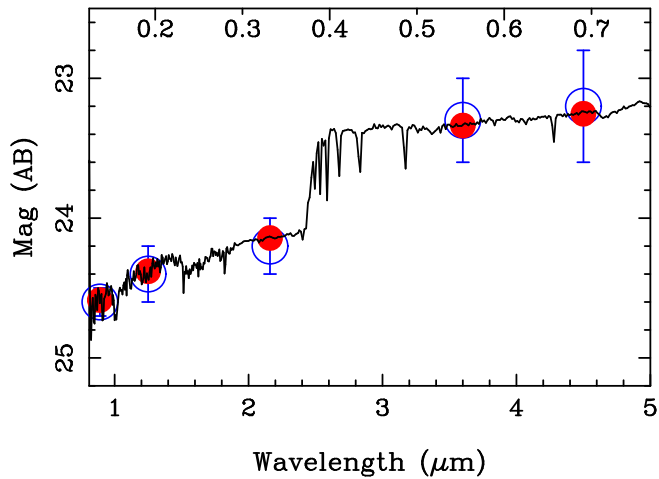}
\caption{\em The best-fit \citet{Bruzual2003} SED for BD38.  The open
  circles with error bars denote the measured magnitudes and the
  closed red circles represent the model SED magnitudes in each
  bandpass and channel.  The upper and lower axes are the rest-frame and observed
  wavelengths, respectively.  Model parameters are  $\tau$ = 100 Myrs,
  $Z_{\sun}$, age = $227^{+27}_{-35}$ Myr, E(B-V) =
  $0.10^{+0.10}_{-0.10}$, and a mass = 43 $\pm$ 13 $10^{9}$ $M_{\sun}$  The magnitudes have
  been dimmed by 0.3 mag to account for weak lensing by the cluster
  potential (\S 3.1).  The clear rest 0.4 $\mu$m Balmer break suggests a dominant stellar population of at least 100 Myr.
\label{sed}}
\end{center}
\end{figure}

We examined three sets of models for the spectral energy distribution
(SED) of BD38: a (1) simple stellar population (SSP), a (2) constant,
or (3) exponentially declining star formation.  All models are poorly
constrained because of having only five independent photometric
measurements and the large errors on the IRAC fluxes.  The best
fitting SSP (minimum mass and age model) is a 0.4 $Z_{\sun}$,
$6.0^{+2.6}_{-0.4}$ Myr population with \textit{E(B-V)} =
$0.26^{+0.04}_{-0.04}$ and a mass of 7.0 $\pm$ 2.1 $\times$ 10$^{9}$
$M_{\sun}$.  Assuming a constant star formation rate, yields a
$Z_{\sun}$, $720^{+470}_{-300}$ Myr population with \textit{E(B-V)} =
$0.13^{+0.07}_{-0.05}$ and a mass of 64 $\pm$ 20 $\times$ $10^{9}$
$M_{\sun}$.  The lowest $\chi^{2}$ model is an exponentially decaying
star formation rate with $\tau$ = 100 Myrs, a $Z_{\sun}$,
$227^{+27}_{-35}$ Myr population with \textit{E(B-V)} =
$0.10^{+0.10}_{-0.10}$ and a mass of 43 $\pm$ 13 $\times$ $10^{9}$
$M_{\sun}$.  All errors are formal 68$\%$ confidence limits for the
input model.  We emphasize that all of these models have competitive
$\chi^{2}$ values and that there are as many parameters ($\tau$, $Z$,
age, \textit{E(B-V)}, and mass) as there are data points.  It is
striking that the most reasonable models (i.e., those with ongoing
star formation) all imply large masses, with mass-to-light ratios
typical of \textit{z} $\approx$ 3 galaxies \citep[e.g.][]{Labbe2005}.

Using the same assumptions as \citet{Eyles2005}, BD38 is at least 2-3
times more massive than the comparably bright ($z_{850}=24.7$)
\textit{z} = 5.78 galaxy SBM03$\#$03, though the stellar ages derived
are similar.  Hence, BD38 is likely the most massive \textit{z} $\sim$
6 \textit{i}-dropout LBG to date.  Interestingly, SBM03$\#$03
possesses strong Ly$\alpha$ emission, is compact (both unlike BD38),
and is 0.6 mag fainter in the 3.6$\mu$ channel than BD38.

This object is unique, representing the tip of the \textit{z} $\sim$ 6
mass function.  BD38 is likely a dusty (\textit{E(B-V)} $\sim$ 0.1),
massive LBG composed of a large current burst of star formation
($SFR_{UV}$ = 40 - 142 $h^{-2}_{0.7} \ M_{\sun} \ \rm{yr^{-1}}$)
together with a slightly older population (6-700 Myr) of 1-6 $\times$
10$^{10}$ $M_{\sun}$ stars.  These results are dependent upon the very
shallow IRAC imaging; much longer integration times are necessary to
confirm this interpretation.  Mid-IR imaging would give much stronger
constraints on possible SEDs.

Though this is only one object, its similarities with other luminous
\textit{z} $\sim$ 3 LBGs give insights into evolution over the
interval 3 $\leq$ \textit{z} $\leq$ 6.  This object has (1) no
noticeable Ly$\alpha$ emission, and (2) is more dust reddenned than
typical \textit{z} $\sim$ 6 objects \citep{Stanway2005,Bouwens2005}.
These traits are typical of the more luminous high $SFR_{UV}$
\textit{z} $\sim$ 3 LBGs \citep{Shapley2003}.  Therefore, even though
the universe was only half as old at \textit{z} = 5.5 (1 Gyr) than at
\textit{z} $\sim$ 3 (2 Gyr), the composition and star forming
properties of extremely luminous objects appear to be essentially
unchanged.

This research has been supported by the NASA grant NAG 5-7697.  We
thank our anonymous referee for helpful comments that improved this
paper.  We are extremely grateful to M. Ando, A. Shapley and
collaborators for the generous use of their data, and Dan Magee for
his adept computer assistance.

\end{document}